\documentclass[doublecol]{epl2} % for 2 columns style with line numbers
\usepackage{graphicx}
\usepackage{amsmath,amssymb,mathrsfs,dcolumn}
\usepackage[colorlinks=true, linkcolor=blue, citecolor=blue, urlcolor=blue, linktoc=page, bookmarks=false, pdfstartview={FitH}, pdfborder={0 0 0.0 [3 3]}]{hyperref}
\hypersetup{pdfborder=0 0 0,colorlinks=true,citecolor=blue,linkcolor=blue}
\title{Recent progress in antiferromagnetic dynamics}

\author{H. Y. Yuan\inst{1}\and Zhe Yuan\inst{2} \thanks{Email: zyuan@bnu.edu.cn} \and Rembert A. Duine \inst{1,3,4} \and X. R. Wang \inst{5,6} \thanks{Email: phxwan@ust.hk}}
\institute{
    \inst{1}Institute for Theoretical Physics, Utrecht University, 3584 CC Utrecht, The Netherlands \\
    \inst{2}Center for Advanced Quantum Studies and Department
of Physics, Beijing Normal University, Beijing 100875, China \\
    \inst{3}Center for Quantum Spintronics, Norwegian University of Science and Technology, NO-7491 Trondheim, Norway\\
    \inst{4}Department of Applied Physics, Eindhoven University of Technology, P.O. Box 513, 5600 MB Eindhoven, The Netherlands\\
    \inst{5} Physics Department, The Hong Kong University of
Science and Technology, Clear Water Bay, Kowloon, Hong Kong\\
    \inst{6} HKUST Shenzhen Research Institute, Shenzhen 518057, China
}


\abstract{Spintronics, since its inception, has mainly focused on ferromagnetic materials for manipulating the spin degree of freedom in addition to the charge degree of freedom, whereas much less attention has been paid to antiferromagnetic materials. Thanks to the advances of micro-nano-fabrication techniques and the electrical control of the N{\'e}el order parameter, antiferromagnetic spintronics is booming as a result of abundant room temperature materials, robustness against external fields and dipolar coupling, and rapid dynamics in the terahertz regime. For the purpose of applications of antiferromagnets, it is essential to have a comprehensive understanding of the antiferromagnetic dynamics at the microscopic level. Here, we first review the general form of equations that govern both antiferromagnetic and ferrimagnetic dynamics. This general form unifies the previous theories in the literature. We also provide a survey for the recent progress related to antiferromagnetic dynamics, including the motion of antiferromagnetic domain walls and skyrmions, the spin pumping and quantum antiferromagnetic spintronics. In particular, open problems in several topics are outlined. Furthermore, we discuss the development of antiferromagnetic quantum magnonics and its potential integration with modern information science and technology.}

\begin{document}

\maketitle

\section{Introduction}
Antiferromagnets (AFMs) are a large class of magnetic materials with magnetic moments on different sublattices of the crystals that compensate each other, which usually leads to vanishingly small net magnetization. Therefore, AFMs are almost immune to an external magnetic field and could, until recently, hardly be manipulated by external means. With the development of the electric writing and reading of the N{\'e}el order in metallic AFMs \cite{MacDonald,ke2008,Jungwirth}, antiferromagnetic spintronics started to attract significant attention. Compared with its widely studied ferromagnetic (FM) counterpart, both metallic and insulating/semiconducting AFMs are more abundant. They are free from cross-talking problems and exhibit faster dynamics at the terahertz (THz) scale in comparison to dynamics at gigahertz time scales for FMs. The ongoing topics of AFM spintronics are quite broad, including spin pumping and spin-transfer torque, spin-orbit-interaction-induced phenomena, (anomalous) spin Hall effect, N{\'e}el-order switching, AFM domain wall/skyrmion dynamics, spin superfluidity, and cavity spintronics, etc., which are reviewed by some very recent articles \cite{Gomonay2014,Jung2016,Baltz2018,Smej2018,Duine2018,Fukami2020}. In this paper, we focus on the recent progress on antiferromagnetic dynamics, which is the basis for the application prospects of AFMs. Specifically, we first review the proper dissipative torques in AFMs, which are essential for a correct and complete theoretical description of AFM dynamics. This is then followed by the dynamics of domain walls (DWs) and skyrmions in AFMs. After a short discussion of spin pumping and magnon transport, we turn to the novel field of quantum antiferromagnetic spintronics, highlighting the quantum nature of magnons and their coupling with cavity photons. This provides the possibility of realizing continuous variable quantum information. Since AFM spintronics is still under development, there are many open questions on all topics that may be interesting for our readership.

\section{Dynamic equations}
The magnetization dynamics for a FM are well described by the Landau-Lifshitz-Gilbert (LLG) equation \cite{Landau1935,Gilbert2004},
\begin{equation}
\partial_t \mathbf{m}= -\mathbf{m} \times  \mathbf{h}_{\mathrm{eff}}
+ \alpha \mathbf{m} \times \partial_t\mathbf{m},
\label{llg}
\end{equation}
where the first and second terms on the right hand side represent the precessional and dissipative torques, respectively, $\mathbf{h}_{\mathrm{eff}}$ is the effective field and $\alpha$ is a phenomenological damping parameter.

The dynamic equations for an AFM had not been settled until a very recent work that unified different approaches in the literature. \textit{Kittel et al.} used the coupled LLG equations for each sublattice of an AFM to describe the AFM resonance in the 1950s \cite{Kittel1951, Keffer1952}. Their approach has been widely adopted in the community and was employed in recent years to study spin-transfer torque, spin wave excitation and DW dynamics \cite{Helen2014,ke2008, Kampfrath2010, Selzer2016}. More recently, \textit{Hals et al.} proposed an alternative set of equations in which the dissipative torques are introduced in a phenomenological way \cite{Hals2011, Tveten2016}.
The resulting equations contain two damping parameters $\alpha_m$ and $\alpha_n$ that characterize the dissipation associated with the motions of magnetization and staggered order, respectively. Without the knowledge of the magnitudes of $\alpha_{m(n)}$, the latter approach was applied to the studies of the AFM dynamics \cite{Tveten2013,Troncoso2015,Shiino2016} by assuming $\alpha_m = \alpha_n$ or $\alpha_m=0$ \cite{Kim2014}. On the other hand, first-principles calculations have demonstrated that $\alpha_m$ is one to three orders of magnitude larger than $\alpha_n$ \cite{ly2017,Mah2018},
which is in sharp contrast to the previous assumptions. Therefore, a rigorous derivation of the proper dissipative torques in antiferromagnetic systems is urgently needed.

\textit{Yuan et al.} considered an $N-$sublattice AFM with sublattice magnetization $\mathbf{m}_1, \mathbf{m}_2,\cdots,\mathbf{m}_N$, and proposed the Lagrangian $\mathcal{L}=\mathcal{T}-\mathcal{U}$, with the kinetic energy $\mathcal{T}=\sum_{i=1}^N \mathbf{A}(\mathbf{m}_i)\cdot  \dot{\mathbf{m}}_i$ and the potential energy $\mathcal{U}$. A Rayleigh dissipation function $\mathcal{R}=(\mathbf{v}\cdot \mathbf{R} \cdot \mathbf{v}^T)/2$ describes the dissipation effect, where $\mathbf{v}=( \dot{\mathbf{m}}_1, \dot{\mathbf{m}}_2,\cdots,\dot{\mathbf{m}}_N)$ and $\mathbf{R}$ is a positive-definite and symmetric $N \times N$ matrix to satisfy the second law of thermodynamics \cite{yuan2019epl}. The magnetization dynamics are then described by the Lagrange equation,
\begin{equation}
\partial_t \mathbf{m}_i= -\mathbf{m}_i \times  \mathbf{h}_i + \mathbf{m}_i \times \left ( \sum_{j=1}^{N} R_{ij}\partial_t \mathbf{m}_j \right ).\label{nsub}
\end{equation}
This is a set of generalized dynamic equations for an $N-$sublattice AFM, featuring the intersublattice dissipation terms $R_{ij}$ ($i\neq j$). For a two-sublattice AFM ($N=2$), we can denote $R_{11}=R_{22}=\alpha$, $R_{12}=R_{21}=\alpha_c$, and simplify eq.~\eqref{nsub} to the form,
\begin{equation}
\begin{aligned}
\partial_t \mathbf{m}_1= -\mathbf{m}_1 \times  \mathbf{h}_1
+ \mathbf{m}_1 \times \left ( \alpha \partial_t \mathbf{m}_1
+ \alpha_c \partial_t \mathbf{m}_2\right ),\\
\partial_t \mathbf{m}_2= -\mathbf{m}_2 \times  \mathbf{h}_2
+ \mathbf{m}_2 \times \left ( \alpha \partial_t \mathbf{m}_2
+ \alpha_c \partial_t \mathbf{m}_1 \right ).
\label{2sub}
\end{aligned}
\end{equation}
Here, the new $\alpha_c$ torque can also be interpreted as the effect of spin pumping between the two sublattices, which gives the constraint $\alpha_c>0$. A similar set of equations was also obtained for a ferrimagnet \cite{Karma2018}.

To investigate the effect of the $\alpha_c-$torque on the AFM dynamics, it is useful to rewrite eq.~(\ref{2sub}) in terms of magnetization $\mathbf{m}  = \mathbf{m}_1 + \mathbf{m}_2$ and staggered order parameter $\mathbf{n}  = \mathbf{m}_1 - \mathbf{m}_2$,
\begin{equation}
\begin{aligned}
&\partial_t \mathbf{m} = -\mathbf{m} \times  \mathbf{h}_m
-\mathbf{n} \times  \mathbf{h}_n + \mathbf{T}_m, \\
&\partial_t \mathbf{n} = -\mathbf{m} \times  \mathbf{h}_n
-\mathbf{n} \times  \mathbf{h}_m + \mathbf{T}_n,  \\
\end{aligned}
\label{llgmn}
\end{equation}
where $\mathbf{h}_m =-\delta \mathcal{U}/\delta \mathbf{m}$ and $\mathbf{h}_n =-\delta \mathcal{U}/\delta \mathbf{n}$ are the effective magnetization field and the N\'{e}el field, respectively.
$\mathbf{T}_m =\alpha_m \mathbf{m} \times \partial_t \mathbf{m}+ \alpha_n \mathbf{n} \times \partial_t \mathbf{n}$ and $\mathbf{T}_n = \alpha_n \mathbf{m} \times \partial_t \mathbf{n}+\alpha_m \mathbf{n} \times \partial_t \mathbf{m}$ are the dissipative torques exerted on $\mathbf{m}$ and $\mathbf{n}$.
Here one has $\alpha_m = (\alpha + \alpha_{c})/2$ and $\alpha_n = (\alpha -\alpha_{c})/2$. The relation $\alpha_n \ll \alpha_m$ obtained by first-principles calculations, which are listed in table \ref{tab1}, can be naturally understood by considering the value of $\alpha_c$ which is smaller but close to the intrinsic damping $\alpha$ in those AFMs.

\begin{table}[htb]
\centering
\caption{First-principles calculation of the resistivity and damping for metallic AFMs. Adapted from Ref. \cite{ly2017}.}
\begin{tabular}{lcccc}
 \hline
            \multicolumn{1}{c} {AFM} &
            \multicolumn{1}{c} {$\mathbf n$} &
            \multicolumn{1}{c} {$\rho$ ($\mu\Omega$ cm)} &
            \multicolumn{1}{c} {$\alpha_n$ ($10^{-3}$) } &
            \multicolumn{1}{c} {$\alpha_m$} \\
\hline
 PtMn  & $a$ axis & 119$\pm$5 & 1.60$\pm$0.02 & 0.49$\pm$0.02 \\
           & $c$ axis & 108$\pm$4 & 0.67$\pm$0.02 & 0.59$\pm$0.02 \\
 IrMn   & $a$ axis & 116$\pm$2 & 10.5$\pm$0.2 & 0.10$\pm$0.01 \\
           & $c$ axis & 116$\pm$2 & 10.2$\pm$0.3 & 0.10$\pm$0.01 \\
 PdMn & $a$ axis & 120$\pm$8 & 0.16$\pm$0.02 & 1.1$\pm$0.10 \\
           & $c$ axis & 121$\pm$8 & 1.30$\pm$0.10 & 1.30$\pm$0.10 \\
 FeMn & $a$ axis & 90$\pm$1 & 0.76$\pm$0.04 & 0.38$\pm$0.01 \\
           & $c$ axis & 91$\pm$1 & 0.82$\pm$0.03 & 0.38$\pm$0.01 \\
 \hline
\end{tabular}
\label{tab1}
\end{table}

Recently, \textit{Simensen et al.} proposed a magnon decay theory induced by the $s-d$ exchange coupling in metallic AFMs and found that $\alpha_n \approx 0, \alpha_m=\pi V^2J^2g^2(\mu)$, where $V$ is the volume of a unit cell, $J$ is the strength of $s-d$ coupling, and $g(\mu)$ is the density of states at the Fermi level $\mu$ \cite{Simen2020}. This is consistent with the conclusions from the rigorous derivation and first-principles calculations that $\alpha_n$ arises from spin-orbit coupling and $\alpha_m$ is attributed to the exchange coupling with $\alpha_n \ll \alpha_m$. For FeMn, the authors find $\alpha_m=0.35$, which is comparable to the first-principles calculation result of 0.38. This highlights the scattering of magnons and conduction electrons as the main source of damping in metallic AMFs. Nevertheless, the justification of $\alpha_m$ and $\alpha_n$ in insulating AFMs from a microscopic theory, as well as first-principles calculation, is still an open question. In general, the effect of $\alpha_m$ will become significant in an AFM with stronger anisotropy \cite{ly2017,yuan2019epl,Simen2020}. Furthermore, \textit{Akosa et al.} first reported the presence of a chiral damping term in noncentrosymmetric ferromagnets \cite{Akosa2016, Jue2016, Akosa2018}. How this chiral damping manifests itself as a dissipative torque in an AFM is not yet known.

\section{Domain wall dynamics} Manipulating DW motion in magnetic nanowires is of particular interest not only for fundamental physics but also for proposed magnetic devices, such as magnetic sensors \cite{Borie2017}, DW logic \cite{Luo2020}, racetrack memory \cite{Parkin2008} and magnetic memristors \cite{Wang2009}. The DW velocity is a key quality determining the usefulness of these devices. In an FM, the DW velocity is mainly influenced by the pinning strength due to the inhomogeneities and disorders in the nanowire \cite{Yuan2014,Chen2019}, and by the Walker breakdown \cite{Walker1974}. While the pinning can be reduced by improving material quality, the breakdown field is an intrinsic limit in ferromagnets, above which the DW velocity is significantly reduced. AFM DWs are free from the Walker breakdown \cite{Helen2016} and can therefore achieve a much larger drift velocity.

\begin{figure}
  \centering
   \includegraphics[width=0.45\textwidth]{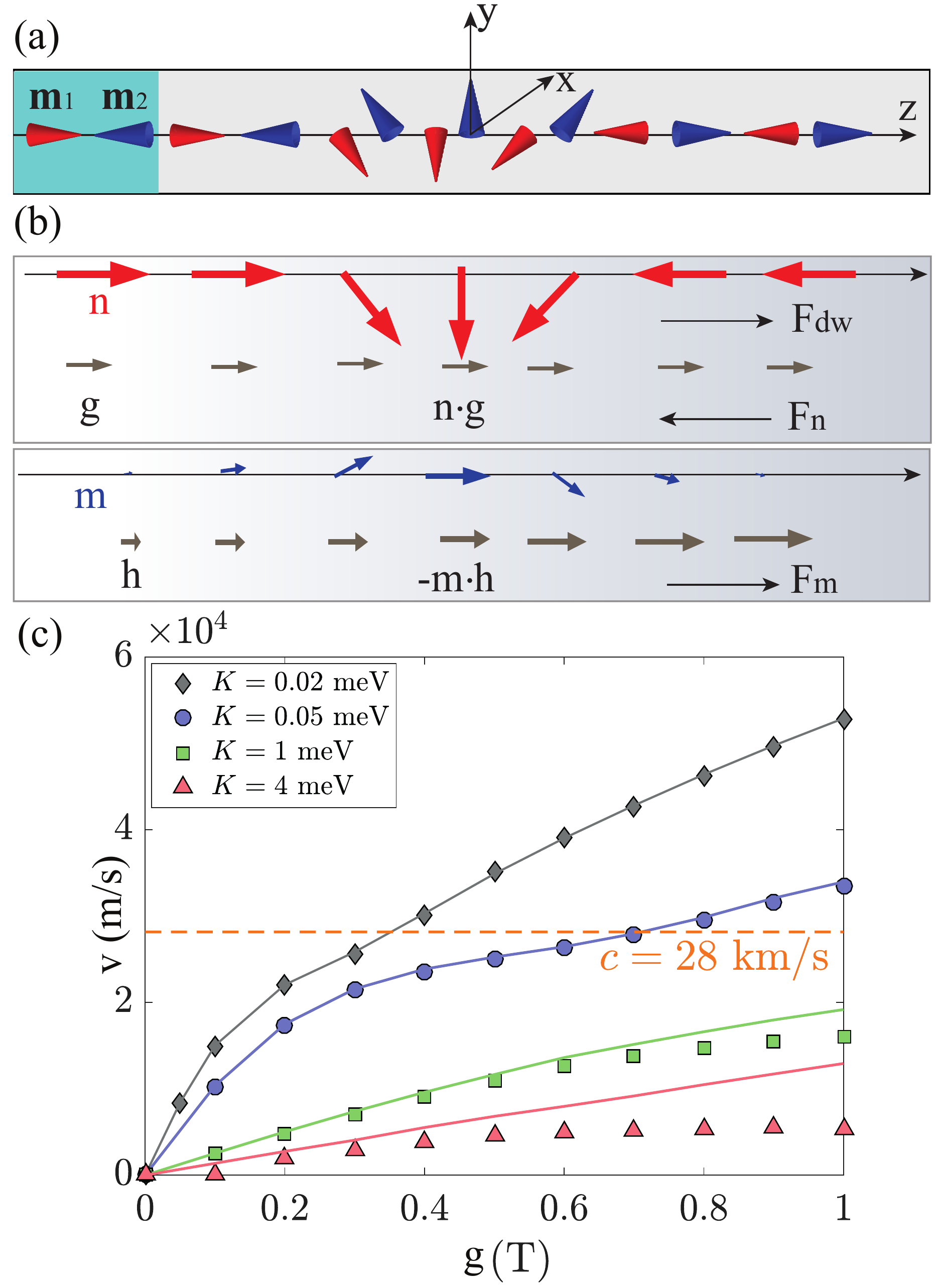}\\
  \caption{(a) Sketch of a two-sublattice AFM
nanowire with a $180^\circ$ DW.
(b) Spatial distribution of staggered order $\mathbf{n}$
(red arrow) and magnetization $\mathbf{m}$
(blue arrow) in a $180^\circ$ DW. (c) DW velocity as a function of the driving field for a nanowire with different anisotropy. Adapted from Ref. \cite{yuan2019atomic,Yuan2018direction}.}\label{direction}
\end{figure}

To give a clear illustration, we consider a two-sublattice AFM [fig.~\ref{direction}(a)] described by the Hamiltonian $\mathcal{H}= \frac{a}{2} \mathbf{m}^2 + \frac{A}{2} (\partial_z\mathbf{n})^2 + L  \mathbf{m} \cdot
 \partial_z \mathbf{n} -\frac{K}{2} n_z^2 - m_zh(z) + n_z g(z)$, where $a$ and $A$ are respectively the homogenous and inhomogeneous exchange coefficients, $K$ is the uniaxial anisotropy, $L$ is the parity breaking term that comes from the permutation symmetry breaking of nonlinear structures, and $h$ and $g$ are respectively the average field over a unit cell and the staggered field. By eliminating the magnetization order in the dynamic equations under the assumption of $|\mathbf{m}| \ll 1$, the equation for the staggered order $\mathbf{n}$ becomes
\begin{equation}
\mathbf{n} \times \left (  -\ddot {\mathbf{n}}+ \alpha_m \dot{\mathbf{h}}_n + a \mathbf{h}_n- \alpha_n a \dot{\mathbf{n}}\right ) =0.
\label{neq}
\end{equation}

For a rigid-body DW propagation, we use the ansatz of $\mathbf{n}=\mathbf{n}(z-z_c)$, where $z_c=vt$ and $v$ are respectively the DW position and velocity. By determining the vector product, $\partial_z \mathbf{n} \times \mathrm{eq}.~(\ref{neq})$, and integrating over the whole space, we obtain the Thiele equation \cite{Thiele1972, Tveten2014,Yuan2018direction},
\begin{equation}
M_{zz} ( \ddot{ z}_c + a \alpha_n \dot{z}_c ) = F_z,
\end{equation}
where $M_{zz}\equiv(1/a)\int (\partial_z\mathbf{n})^2dz=2/(a \Delta)$. The effective driving force on the right-hand-side contains three terms, i.e., $F_z=F_m+F_n+F_{DW}$ (see fig.~\ref{direction}(b)) with
\begin{equation}
\begin{aligned}
&F_m =\frac{ L}{a\Delta}\int({\partial_z}h)\sin^2\theta\ dz \ ,\\
&F_n=-\frac{1}{\Delta}\int g \sin^2\theta \ dz \ ,\\
&F_{DW}=\frac{1}{2A}\int ( \partial_z h^2)\sin^2\theta dz \ ,\\
\end{aligned}
\label{threeF}
\end{equation}
where $\theta$ is the angle between $\mathbf{n}$ and the $z$ axis. For a steady DW motion $\ddot{z}_c =0$, the DW velocity is
\begin{equation}
v = (\Delta /2\alpha_n )\left(F_m+F_n+ F_{dw} \right).
\label{th_v}
\end{equation}
Here, the staggered field $g$ can be generated by spin-orbit torque \cite{Zelezny2014,Shiino2016} and is the most efficient DW driven force with a high speed, since the normal inhomogeneous field $ h(z)$ is limited by the spin-flop field of the system \cite{Yuan2018direction}. It is straightforward to solve the steady DW velocity driven by a staggered field. From eq.~(\ref{th_v}), we have $v=c g\Delta_0/\sqrt{g^2 \Delta_0^2 + \alpha_n^2 c^2}$, in which the relativistic behavior of the DW width $\Delta = \Delta_0\sqrt{1-v^2/c^2}$ is used, where $\Delta_0$ is the static DW width and $c$ is the magnon speed. This result indicates that the DW velocity increases with the driving force and is saturated at the magnon speed $c$.

Surprisingly, \textit{Yang et al.} found that the DW velocity can exceed the relativistic limit \cite{yuan2019atomic}, as shown in fig. \ref{direction}(c). This occurs for the atomically thin DWs, becoming unstable as the drift velocity approaches the limit. The strongly emitted spin waves then accumulate at the tail of the DW, resembling Cherenkov radiation in electrodynamics. The tail modifies the atomic DW profile, resulting in a further increase in the effective DW width. Thus, the DW speed can become larger than the relativistic limit. We still need a microscopic theory to correctly model the interplay of spin waves and AFM DWs, and hence to have the capability of predicting the supermagnonic motion of DWs.

In addition to the spin-orbit field, AFM DWs can be driven by an electric current via spin transfer torque \cite{Hals2011}, an inhomogeneous magnetic field \cite{Yuan2018direction}, spin waves \cite{Tveten2014,Lan2018}, microwaves \cite{Pan2018} and a thermal gradient \cite{Selzer2016}. The AFM domain and/or DWs in experimental samples can be imaged by the optical methods, X-ray techniques and by a spin-polarized tunneling microscope \cite{Cheong2020}. Unfortunately, it is still challenging to combine these imaging techniques with various driving knobs (electric current, field gradient, magnon current, etc.) to study the dependence of DW velocity on the driving strength.

\section{Skyrmion dynamics}
In addition to the DWs, the skyrmion is another topological nontrivial structure with noncollinear magnetization, which is a suitable candidate for studying topological physics and is also proposed as an information carrier in devices. AFM skyrmions are insensitive to the external magnetic field and do not have the undesired skyrmion Hall effect as the FM skyrmions do. Various techniques to nucleate a single AFM skyrmion have been theoretically proposed, such as injection of a vertical spin-polarized current, conversion from a pair of AFM DWs \cite{Zhang2016}, and illumination with a short laser pulse \cite{Kho2019}. The skyrmion lattice may even be stabilized in a frustrated AFM material without Dzyaloshinskii-Moriya interaction \cite{Okubo2012,Rosales2015}. Nevertheless, the generation and propagation of AFM skyrmions is still very challenging in experimentation. Recently, \textit{Dohi et al.} and \textit{Legrand et al.} demonstrated room-temperature AFM skyrmions in a synthetic antiferromagnet (SAF) \cite{Dohi2019,Leg2020}.

Concerning the dynamics of AFM skyrmions, \textit{Barker et al.} first showed that the skyrmion motion was strictly along the electric current (without the skyrmion Hall effect) in an AFM \cite{Barker2016}. Owing to the technical difficulty of engineering the skyrmion in the conventional bipartite AFMs, \textit{Zhang et al.} proposed to use SAF for skyrmion propagation, where the Magnus forces acting on a pair of skyrmions in each FM layer cancel each other out. This idea was later demonstrated in an SAF hosting skyrmion bubbles driven by a small electric current \cite{Dohi2019}. \textit{Daniels et al.} found that the Hall physics is quite different when the skyrmion was driven by a chiral spin current \cite{Daniels2019}. This may provide an alternate route to manipulate AFM skyrmions in an insulating material. Moreover, as mentioned in the previous section, the disorder and geometrical pinning determines the current density threshold of DW motion in a magnetic nanostructure. Similar effects exist for skyrmions, but few studies have considered the influence of disorder on the AFM skyrmion dynamics. In principle, the Magnus force can help an FM skyrmion to avoid individual impurities, while the straight motion of AFM skyrmions may experience more hindering effects from the pinning sites \cite{Liang2019}. The issue may become more complicated in a granular film \cite{Gong}. In addition, the nonlocal damping due to the noncollinear magnetization may become significant when the skyrmions are small and when a large magnetization gradient is present, which will be discussed in the next section.

\section{Spin current pumping and magnon transport} When magnetization precesses, the spin-dependent chemical potential is varying in time, resulting in a driven spin current. However, such a spin current has an observable consequence only if the FM system has broken translational symmetry, such as an interface \cite{Yaroslav2002} or noncollinear magnetization \cite{YuanZ2014}. The spin current is then absorbed up by an electron reservoir in contact with the magnetic material and enhances the magnetic damping. In inhomogeneous magnetic systems, the pumped spin current can also be absorbed by the neighboring spins and gives rise to an additional spin torque, which hence influences the local magnetization dynamics.

\textit{Cheng et al.} examined the scattering of electrons at the interface of a collinear AFM and a normal metal \cite{Cheng2014,Takei2014} and found that both the motion of magnetization and staggered order parameter could pump a spin current of
\begin{equation}
\begin{aligned}
I_s=G_r(\mathbf{n}\times \mathbf{\dot{n}}+ \mathbf{m}\times \mathbf{\dot{m}}),\\
I_{ss}=G_r(\mathbf{m}\times \mathbf{\dot{n}}+ \mathbf{n}\times \mathbf{\dot{m}}),
\end{aligned}
\label{spinpump}
\end{equation}
where $G_r$ is the spin mixing conductance of the interface. It is straightforward to rewrite the total spin current in eq. (\ref{spinpump}) as $I=G_r \mathbf{m}_1\times \mathbf{\dot{m}}_1 + G_r\mathbf{m}_2\times \mathbf{\dot{m}}_2$, which assumes no intersublattice contributions of $\mathbf{m}_1\times \mathbf{\dot{m}}_2$ and $\mathbf{m}_2\times \mathbf{\dot{m}}_1$. Later, the existence of these crossing terms was justified by \textit{Kamra et al.} \cite{Kamra2017} and was also supported by first-principles calculations \cite{ly2017} of the damping parameters listed in table \ref{tab1}. Recently, \textit{Li et al.} reported the detection of the spin current generated by an AFM ($\mathrm{Cr_2O_3}$) by converting it to an electric signal in an adjacent heavy metal layer through the inverse spin Hall effect \cite{Li2020}. For the intersublattice spin pumping that is predicted to be larger than the damping of the staggered order, experimental verification is yet to be carried out. Interestingly, \textit{Shen} reported an intrinsic mechanism for the generation of magnon spin current from the symmetry breaking in AFMs caused by the dipolar interaction \cite{Shen2020}.

In addition to the spin current pumping, an AFM can absorb spin currents emitted from various sources such as a thermal gradient \cite{Lin2016}, strongly coupled FM resonance \cite{Wang2014}, and neighboring heavy metals with spin-orbit interaction \cite{Takei2014}. Under this circumstance, the transport and dynamic properties of the AFM may be artificially manipulated. \textit{Takei et al.} theoretically mapped the AFM dynamics to the motion of superfluid and predicted the dissipationless spin current transport in an AFM sandwiched by normal metals \cite{Takei2014}. Soon after, \textit{Yuan et al.} reported experimental signatures of the spin superfluid by measuring the nonlocal spin transport via thermal injection \cite{Yuan2018}. The propagation of the spin signal with distance can be effectively described by a superfluid model. In contrast, \textit{Lebrun et al.} argued that the long-distance transport can also be explained by a diffusive model, which rules out the superfluidity \cite{Lebrun2018}. Reconciling these experiments and theories in a consistent way requires further investigations \cite{Sonin2019}.

For an AFM material hosting noncollinear spin textures, such as DWs, vortices, and skyrmions, one spin may feel unbalanced spin current generated by its neighboring spins, and thus a net spin torque is exerted to influence its dynamics. This is the so-called nonlocal dissipative torque, in analogy with its FM counterpart. The form of the torque can be obtained by including the Rayleigh dissipation terms $R_{nl}=\eta(\partial_t\nabla \mathbf{m}_1)^2 +\eta(\partial_t\nabla \mathbf{m}_2)^2 + \eta_c(\partial_t\nabla \mathbf{m}_1)\cdot(\partial_t\nabla \mathbf{m}_2)$ in the Lagrange equation, whereas its influence on the dynamics of AFM textures is yet to be studied in detail.

\section{Quantum antiferromagnetic spintronics}

In the above sections, the spins we discussed are essentially classical angular momenta. Inspired by the swift development of quantum science and technology, it would be interesting and useful to consider the quantum states in AFM materials and the related excitations. Many related questions arise naturally if one considers AFM textures as a quantum state. For example, what are the transport and dynamic properties of quantum AFM textures? Do quantum AFM textures and their elementary excitations have any intriguing physics or potential applications in quantum information processing? These topics belong to the nascent field of quantum AFM spintronics.

\subsection{Quantum magnetic state} The AFM exchange interaction in quantum systems usually leads to a ground state with strong quantum fluctuations, such as a quantum spin liquid. \textit{Hirobe et al.} first showed that a quantum AFM chain carries a spin current, which could be directly observed by the spin Seebeck effect \cite{Hirobe2017}. Differing from the spin currents carried by electrons or magnons, this spin current in the quantum AFM chain is attributed to the spinons. Owing to its quantum nature, this spin current can be used as a probe for novel quantum states and quantum phase transition in exotic quantum magnets \cite{Camilo2018,Han2020}.

To examine noncollinear AFM textures, \textit{Yuan et al.} used a strong local field to suppress the spin fluctuation of spins at the boundary in a transverse spin-1/2 AFM chain and found that the ground state resembles a DW profile \cite{yuanqm}. Such a quantum DW is driven by an external field and spin-orbit field as a classical DW, but the spins inside the DW are entangled. This entanglement not only provides a characterization of the phase transition of the system from a DW to a domain state but also provides a platform to study quantum information. In principle, the essential physics can be generalized to quantum skyrmions in two dimensions, although this is yet to be accomplished for an AFM. The quantum skyrmion in an FM spin-1 square lattice was recently reported \cite{Wieser2017}.

\subsection{Cavity AFM spintronics}
Cavity AFM spintronics manipulates the interplay of AFM magnons with the cavity photons, aiming to realize coherent information transfer between magnetic and photonic systems. The conventional theory based on classical electrodynamics is sufficient to recover the avoided crossings spectrum near the resonance \cite{Man1972,Bose1975,Xiao2019}, while the quantum-mechanical picture is necessary to consider the quantum correlations among magnons and photons \cite{yuan2017,yuan2020}. This coherent magnon-photon system is an ideal platform for coding and processing quantum information, such as quantum transducer and quantum memory \cite{Parvini2020}.

\begin{figure}
  \centering
  \includegraphics[width=0.45\textwidth]{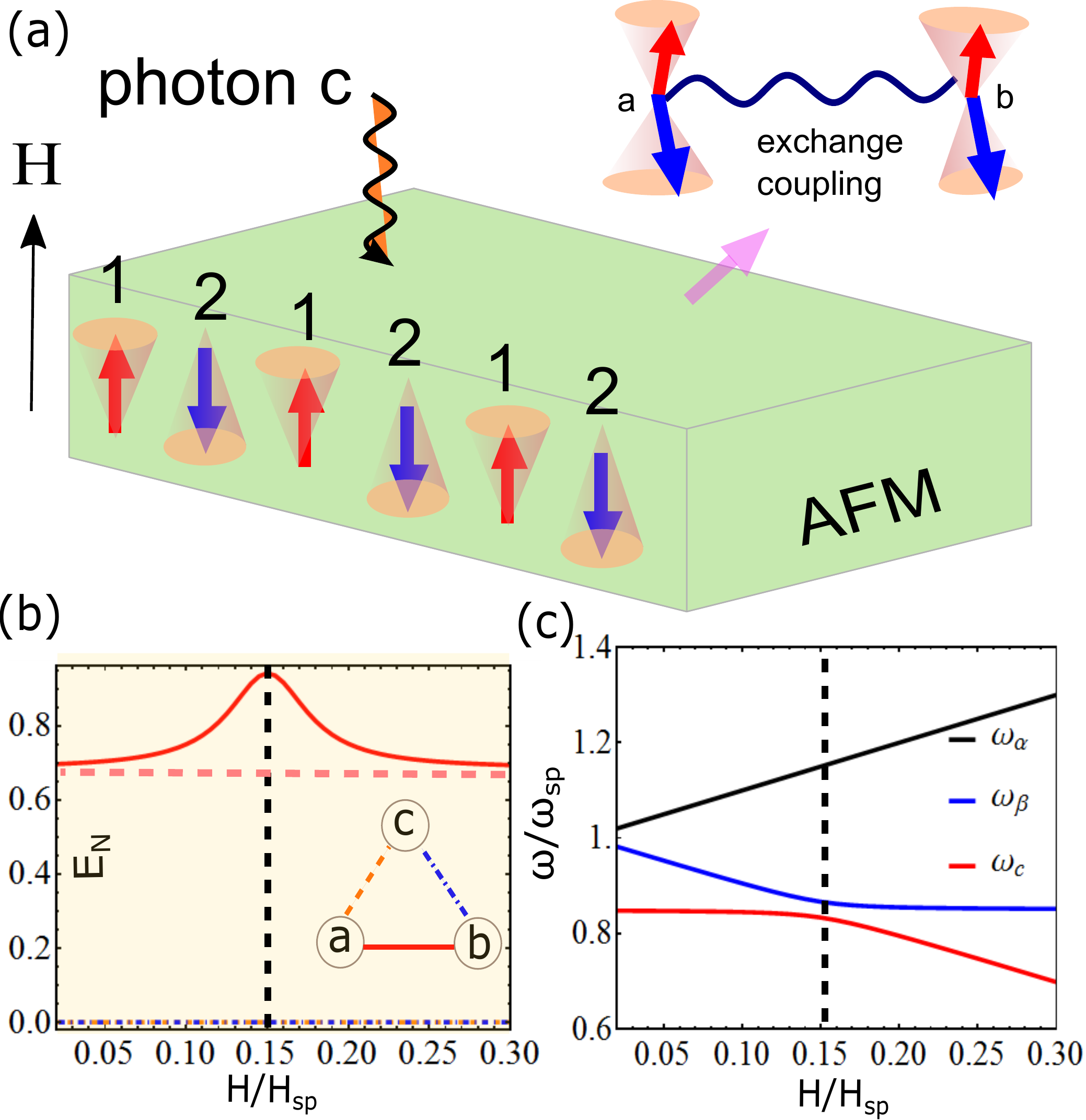}\\
  \caption{(a) Schematic illustration of magnon-photon interaction in AFM materials. (b) Steady entanglement among modes $a$, $b$ and $c$ as a function of the external field. The horizontal dashed line is the entanglement of two magnons without light. (c) Coupling spectrum of a hybrid AFM-light system. Adapted from Ref.~\cite{yuan2020}.}\label{mpen}
\end{figure}

In general, light with a definite circular polarization can interact with a magnet through angular momentum transfer. Due to the opposite magnetization of the two sublattices in an AFM, creating a magnon after annihilating a photon with the angular momentum $\hbar$ is described by the magnon operator $a$ on sublattice 1 and $b^\dagger$ on sublattice 2, respectively, as shown in fig.~\ref{mpen}(a), instead of $a^\dagger$ and $b^\dagger$. This leads to the following Hamiltonian of a hybrid magnon-photon system, \cite{yuan2020},
 \begin{eqnarray}
\mathcal{H}&=& \omega_a a^\dagger a + \omega_b b^\dagger b +g_{ab}(a^\dagger
b^\dagger + a b) + \omega_c c^\dagger c  \nonumber\\
&&+ g_{ac} \left ( a^ \dagger c^\dagger + a c \right)
+ g_{bc} \left (  b^\dagger c + b c^\dagger \right),
\label{FimHm}
\end{eqnarray}
where $a^\dagger$ ($a$) and $b^\dagger$ ($b$) are the creation (annihilation) operators of magnons on sublattice 1 and sublattice 2, respectively. $c$ and $c^\dagger$ are the creation and annihilation operators of photons.
$g_{ab}$, $g_{ac}$ and $g_{bc}$ are their mutual couplings. The sublattice permutation symmetry of an AFM implies $g_{ac}=g_{bc}$.

Using the standard techniques for continuous variable quantum information \cite{yuan2020}, one can quantify the entanglement of modes $a$, $b$, and $c$ by log-negativity ($E_N$), as shown in fig.~\ref{mpen}(b). Without photons, the two types of magnons are already entangled with $E_N=0.69$, which can be readily understood by the squeezing nature of the magnons resulting from their parametric-type coupling. This was also reported by \textit{Kamra et al.} \cite{Kamra2019} and \textit{Mousolou et al.} \cite{Mou2020}, where the entanglement was quantified by the entanglement entropy. In the presence of photons, the magnon-magnon entanglement becomes even stronger and the maximum enhancement appears at the resonance, where the acoustic magnon frequency is equal to the cavity frequency; see fig.~\ref{mpen}(c). This enhanced entanglement is because of the magnon cooling mediated by the cavity photons, and the theory can be straightforwardly generalized to a two-sublattice ferrimagnetic system with $g_{ac} \neq g_{bc}$.

Experimentally, the spectrum of strong magnon-photon coupling was first observed in a crystal of di(phenyl)-(2,4,6-trinitrophenyl)iminoazanium (DPPH) \cite{Merg2017}. \textit{Everts et al.} later reported the ultrastrong coupling between photon and magnon excited in rare-earth ion spins ($\mathrm{GdVO_4}$) \cite{Everts2020} with a level repulsion spectrum. A direct coupling between a terahertz wave and AFM magnon was realized in AFMs ($\mathrm{TmFeO_3}$ and $\mathrm{Fe_2Mo_3O_8}$) with a clear beating pattern near the resonance \cite{Gris2018,Shi2020}. All of these observations require cryogenic conditions around several Kelvin to maintain the AFM order and the magnon excitation. The experimental verification of the magnon-magnon entanglement is still missing, where quantum optical techniques such as homodyne measurement may be engaged.

%\begin{figure}
%  \centering
%  \includegraphics[width=0.4\textwidth]{fig3_level_rep.pdf}\\
%  \caption{Transmission of a hybrid AFM-cavity system in the plane of the external field and wave frequency. The clear avoided crossings spectrum indicates the ultrastrong coupling of AFM magnon and photon. Adapted from Ref. \cite{Everts2020}.}\label{mpc}
%\end{figure}

%\begin{table}
%\caption{List of the typical experiments on AFM magnon-photon coupling. $g_c$ is the maximum coupling strength of the cavity mode and magnon mode.}
%\begin{tabular}{l|c|c|c}
 %            \hline
%           Material & resonator type & $g_c$ (GHz)&T (K)\\
%             \hline
%            DPPH \cite{Merg2017}  & coplanar  & 0.039 & $< 0.05$\\
%            $\mathrm{GdVO_4}$ \cite{Everts2020}  & loop-gap  & 1.75 & $\leq 4$\\
%            $\mathrm{TmFeO_3}$ \cite{Gris2018}  & THz wave & $\sim$26 & $< 90$\\
%            $\mathrm{Fe_2Mo_3O_8}$ \cite{Shi2020}  & THz wave & 0.1 & $\leq 4$\\
%             \hline
%           \end{tabular}
%\label{tab2}
%\end{table}

Until now, we have focused on the magnon excitation in a single magnet and their coupling with electromagnetic waves. It is also interesting to consider the coupling of multiple magnets with the assistance of photons. \textit{Johansen et al.} found that AFM and FM were coupled inside a microwave cavity, where the coupling strength proportional to their respective coupling with the microwave can become comparable with that of two FMs \cite{Johansen2018}. In principle, two AFMs can also be entangled in the same manner, although this remains to be verified. This nonlocal quantum coupling between two magnets may be useful in bipartite continuous variable quantum information processing, such as the asymmetric quantum steering of the two macroscopic magnets.

%Furthermore, one can place a periodic structure of AFM inside a cavity, and examine its nontrivial topological %properties \cite{Pirm2018}.

\section{Conclusions and Outlook}
%Quality Control Editor: Please use a consistent capitalization and punctuation format for section headings throughout the manuscript. Some journals request a specific style, so please review the journal's guidelines.
In conclusion, we have reviewed the recent progress on AFM dynamics. In addition to the open questions outlined in each section, we would like to highlight the promising but largely unexplored direction of antiferromagnetic quantum magnonics that involves creation, manipulation, and applications of the quantum nature of magnons rather than the classical nature of spin waves. It includes the generation and manipulation of macroscopic quantum states of magnets such as spin superfluidity, Bose-Einstein condensation, etc. It has recently been developed to integrate the AFM system with a photonic system and to realize multifunctional quantum information processing. In terms of quantum information and quantum communication, it would be important to generate important magnon quantum states such as the squeezed state, the entangled state, and the single magnon state. These are indispensable resources for quantum information. This is a highly interdisciplinary field involving spintronics, quantum optics and quantum information, and thus may potentially expand the horizon of AFM spintronics. Furthermore, the bipartite nature of the magnonic excitation in a two-sublattice AFM obeys the parity-time symmetry and provides a generic platform to study non-Hermitian physics, such as the exceptional points, positive Gilbert damping, and non-Hermitian topological phases \cite{Ra2017}.

\acknowledgments

This project received funding from the European Research Council (ERC) under the European Union Horizon 2020 research and innovation programme (grant agreement No. 725509). RD is member of the D-ITP consortium, a program of the Netherlands Organisation for Scientific Research (NWO) that is funded by the Dutch Ministry of Education, Culture and Science (OCW). ZY was supported by the National Natural Science Foundation of China (Grant No. 61774018). XRW was supported by the National Natural Science Foundation of China (Grant Nos.~11774296 and 11974296) and Hong Kong RGC (Grant Nos.~16301518 and 16301619).


\begin{thebibliography}{0}

\bibitem{MacDonald}
  \Name{Duine R. A., Haney P. M., Nunez A. S. \and MacDonald A. H.}
  \REVIEW{Phys. Rev. B}{75}{2007}{014433};
  \Name{Wei Z. \textit{et al.}}
  \REVIEW{Phys. Rev. Lett.}{98}{2007}{116603};
  \Name{Haney P. M. \and MacDonald A. H.}
  \REVIEW{Phys. Rev. Lett.}{100}{2008}{196801}.

\bibitem{ke2008}
  \Name{Xu Y., Wang S. \and Xia K.}
  \REVIEW{Phys. Rev. Lett.}{100}{2008}{226602}.

\bibitem{Jungwirth}
  \Name{Wadley P. \textit{et al.}}
  \REVIEW{Science}{351}{2016}{587}.

\bibitem{Baltz2018}
\Name{Baltz V. \textit{et al.}}
\REVIEW{Rev. Mod. Phys.}{90}{2018}{015005}.


\bibitem{Jung2016}
\Name{Jungwirth T., Marti X., Wadley P. \and Wunderlich J.}
\REVIEW{Nat. Nanotech.}{11}{2016}{231}.

\bibitem{Smej2018}
\Name{\v{S}mejkal L., Mokrousov Y., Yan B. \and MacDonald A.H.}
\REVIEW{Nat. Phys.}{14}{2018}{242}.

\bibitem{Gomonay2014}
  \Name{Gomonaya E. V.\and Loktev V. M.}
  \REVIEW{Low. Temp. Phys.}{40}{2014}{17}.

\bibitem{Duine2018}
\Name{Duine, R. A., Lee K.-J., Parkin S. S. P. \and Stiles M. D.}
\REVIEW{Nat. Phys.}{14}{2018}{7}.

\bibitem{Fukami2020}
\Name{Fukami S., Lorenz V. O. \and Gomonay O.}
\REVIEW{J. Appl. Phys.}{128}{2020}{070401}.

\bibitem{Landau1935}
  \Name{Landau L. \and Lifshits E.}
  \REVIEW{Phys. Zeitsch. der Sow.}{8}{1935}{153}.

\bibitem{Gilbert2004}
  \Name{Gilbert T. L.}
  \REVIEW{IEEE Trans. Magn.}{40}{2004}{3443}.

\bibitem{Kittel1951}
  \Name{Kittel C.}
  \REVIEW{Phys. Rev.}{82}{1951}{565}.

\bibitem{Keffer1952}
  \Name{Keffer F. \and Kittel C.}
  \REVIEW{Phys. Rev.}{85}{1952}{329}.

\bibitem{Helen2014}
  \Name{Gomonay E. V. \and Loktev V. M.}
  \REVIEW{Low Temp. Phys.}{40}{2014}{17}.

\bibitem{Selzer2016}
  \Name{Selzer S., Atxitia U., Ritzmann U., Hinzke D. \and Nowak U.}
  \REVIEW{Phys. Rev. Lett.}{117}{2016}{107201}.

\bibitem{Kampfrath2010}
  \Name{Kampfrath T. \textit{et al.}}
  \REVIEW{Nat. Photon.}{5}{2010}{31}.

%\bibitem{Atxitia2017}
%  \Name{Atxitia U., Hinzke D. \and Nowak U.}
%  \REVIEW{J. Phys. D: Appl. Phys.}{50}{2017}{033003}.

\bibitem{Tveten2016}
  \Name{Tveten E. G., Muller T., Linder J. \and Brataas A.}

\bibitem{Hals2011}
  \Name{Hals K. M. D., Tserkovnyak Y. \and Brataas A. }
  \REVIEW{Phys. Rev. Lett.}{106}{2011}{107206}.





\bibitem{Troncoso2015}
  \Name{Trtoncoso R. E., Ulloa C., Pesce F. \and Nunez A. S.}
  \REVIEW{Phys. Rev. B}{92}{2015}{224424}.

\bibitem{Shiino2016}
  \Name{Shiino T. \textit{et al.}}
  \REVIEW{Phys. Rev. Lett.}{117}{2016}{087203}.


%\bibitem{experiments}
%  \Name{Ono T., Miyajima H., Shigeto K., Mibu K., Hosoito N. \and Shinjo T.}
%  \REVIEW{Science}{284}{1999}{468};
%  \Name{Beach G. S. D., Knutson C., Nistor C., Tsoi M. \and Erskine J. L.}
%  \REVIEW{Phys. Rev. Lett.}{97}{2006}{057203};
%  \Name{Hayashi M., Thomas L., Bazaliy Y. B., Rettner C., Moriya R., Jiang X. \and Parkin S. S. P.}
%  \REVIEW{Phys. Rev. Lett.}{96}{2006}{197207};
%  \Name{Kl\"{a}ui M., Jubert P. O., Allenspach R., Bischof A., Bland J. A. C., Faini G., R\"{u}diger U., Vaz C. A. F., Vila L. \and Vouille C.}
%  \REVIEW{Phys. Rev. Lett.}{95}{2005}{026601}.
%
%\bibitem{theory}
%  \Name{Schryer N. L. \and Walker L. R.}
%  \REVIEW{J. Appl. Phys.}{45}{1974}{5406};
%  \Name{Tatara G. \and Kohno H.}
%  \REVIEW{Phys. Rev. Lett.}{92}{2004}{086601};
%  \Name{Zhang S. \and Li Z.}
%  \REVIEW{Phys. Rev. Lett.}{92}{2004}{127204};
%  \Name{Barnes S. E. \and Maekawa S.}
%  \REVIEW{Phys. Rev. Lett.}{95}{2005}{107204};
%  \Name{Wang X. R., Yan P., Lu J. \and He C.}
%  \REVIEW{Ann. Phys. (N. Y.)}{324}{2009}{1815};
%  \Name{Wang X. R., Yan P. \and Lu J.}
%  \REVIEW{Europhys. Lett.}{86}{2009}{67001};
%  \Name{Yuan H. Y. \and Wang X. R.}
%  \REVIEW{Phys. Rev. B}{92}{2015}{054419};
%  \Name{Yuan H. Y., Yuan Zhe, Xia Ke \and Wang X. R.}
%  \REVIEW{Phys. Rev. B}{94}{2016}{064415}.





%  \REVIEW{Phys. Rev. B}{93}{2016}{104408}.

\bibitem{Tveten2013}
  \Name{Tveten E. G., Qaiumzadeh A., Tretiakov O. A. \and Brataas A.}
  \REVIEW{Phys. Rev. Lett.}{110}{2013}{127208}.

%\bibitem{Takei2014}
%  \Name{Takei S., Halperin B. I., Yacoby A. \and Tserkovnyak Y.}
%  \REVIEW{Phys. Rev. B}{90}{2014}{094408}.

\bibitem{Kim2014}
  \Name{Kim S. K., Tserkovnyak Y. \and Tchernyshyov O.}
  \REVIEW{Phys. Rev. B} {90}{2014}{104406}.

\bibitem{ly2017}
  \Name{Liu Q., Yuan H. Y., Xia K. \and Yuan Zhe}
  \REVIEW{Phys. Rev. Mater.}{1}{2017}{061401(R)}.

\bibitem{Mah2018}
  \Name{Mahfouzi F. \and Kioussis N.}
  \REVIEW{Phys. Rev. B}{98}{2018}{220410(R)}.

\bibitem{yuan2019epl}
  \Name{Yuan H. Y., Liu Q., Xia K., Yuan Z. \and Wang X. R.}
  \REVIEW{EPL}{126}{2019}{67006}.

\bibitem{Karma2018}
  \Name{Kamra A., Troncoso R., Belzig W. \and Brataas A.}
  \REVIEW{Phys. Rev. B}{98}{2018}{184402} (2018).

\bibitem{Simen2020}
  \Name{Simensen H. T., Kamra A., Troncoso R. E. \and Brataas A.}
  \REVIEW{Phys. Rev. B}{101}{2020}{020403}.

\bibitem{Akosa2016}
    \Name{Akosa C.A., Miron I.M., Gaudin G. \and Manchon A.}
    \REVIEW{Phys. Rev. B}{93}{2016}{214429}.

\bibitem{Jue2016}
    \Name{ Ju\'{e} E. \textit{et al.}}
    \REVIEW{Nat. Mater.}{15}{2016}{272}.

\bibitem{Akosa2018}
    \Name{Akosa C. A., Takeuchi A., Yuan Z. \and Tatara G.}
    \REVIEW{Phys. Rev. B}{98}{2018}{184424}.

%\bibitem{Kim2018}
%    \Name{Kim K.-W., Lee H.-W., Lee K.-J., Everschor-Sitte K.,
%Gomonay O., \and Sinova J.}

%\bibitem{Gomonay2012}
%  \Name{Gomonay H. V., Kunitsyn R. V. \and Loktev V. M.}
%  \REVIEW{Phys. Rev. B}{85}{2012}{134446}.
% domain wall sectoin
\bibitem{Borie2017}
    \Name{Borie B. \textit{et al.}}
\REVIEW{Phys. Rev. Appl.}{8}{2017}{044004}.

\bibitem{Luo2020}
    \Name{Luo Z. \textit{et al.}}
\REVIEW{Nature}{579}{2020}{214}.

\bibitem{Parkin2008}
    \Name{ Parkin S. S. P., Hayashi M. \and Thomas L.}
\REVIEW{Nature}{320}{2008}{190}.

\bibitem{Wang2009}
   \Name{Wang X. B. \textit{et al.}}
\REVIEW{IEEE Electron Device Lett.}{30}{2009}{294}.

\bibitem{Yuan2014}
  \Name{Yuan H.Y. \and Wang X.R.}
  \REVIEW{Phys. Rev. B}{89}{2014}{054423}.

\bibitem{Chen2019}
  \Name{Chen Z. Y., Qin M. H. \and Liu J.-M.}
  \REVIEW{Phys. Rev. B}{100}{2019}{020402}.

\bibitem{Walker1974}
\Name{ Schryer L. \and Walker L.R. }
\REVIEW{J. of Appl. Phys.}{45}{1974}{5406}.

\bibitem{Helen2016}
  \Name{Gomonay O., Jungwirth T. \and Sinova J.}
  \REVIEW{Phys. Rev. Lett.}{117}{2016}{017202}.

\bibitem{Thiele1972}
 \Name{Thiele A. A.}
  \REVIEW{Phys. Rev. Lett.}{30}{1972}{230}.

\bibitem{Tveten2014}
  \Name{Tveten E. G., Oqiumzadeh A. \and Brataas A.}
  \REVIEW{Phys. Rev. Lett.}{112}{2014}{147204}.

\bibitem{Yuan2018direction}
  \Name{Yuan H.Y., Wang W., Yung M.-H. \and Wang X.R.}
  \REVIEW{Phys. Rev. B}{97}{2018}{214434}.

\bibitem{Zelezny2014}
  \Name{\v{Z}elezn\'{y} J. \textit{et al.}}
  \REVIEW{Phys. Rev. Lett.}{113}{2014}{157201}.

\bibitem{yuan2019atomic}
  \Name{Yang H., Yuan H.Y., Yan M., Zhang H. W. \and Yan P.}
  \REVIEW{Phys. Rev. B}{100}{2019}{024407}.

\bibitem{Lan2018}
  \Name{Weichao Yu, Jin Lan \and Jiang Xiao }
  \REVIEW{Phys. Rev. B}{98}{2018}{144422}.

 \bibitem{Pan2018}
  \Name{Pan K., Xing L., Yuan H. Y. \and Wang W.}
  \REVIEW{Phys. Rev. B}{97}{2018}{184418}.


\bibitem{Cheong2020}
\Name{Cheong S.-W., Fiebig M., Wu W., Chapon L. \and Kiryukhin V.}
\REVIEW{npj Quantum Materials}{5}{2020}{3}.

\bibitem{Zhang2016}
    \Name{Zhang X., Zhou Y. \and Ezawa M.}
\REVIEW{Sci. Rep.}{6}{2016}{24795}.

\bibitem{Kho2019}
\Name{Khoshlahni R., Qaiumzadeh A., Bergman A. \and Brataas A.}
\REVIEW{Phys. Rev. B}{99}{2019}{054423}.

\bibitem{Okubo2012}
    \Name{Okubo T., Chung S. \and Kawamura H.}
\REVIEW{Phys. Rev. Lett.}{108}{2012}{017206}.

\bibitem{Rosales2015}
    \Name{Rosales H.D., Cabra D.C. \and Pujol P.}
\REVIEW{Phys. Rev. B}{92}{2015}{214439}.

\bibitem{Dohi2019}
    \Name{Dohi T., DuttaGupta S., Fukami S. \and Ohno H.}
\REVIEW{Nat. Commun.}{10}{2019}{5153}.

\bibitem{Leg2020}
    \Name{Legrand W. \textit{et al.}}
\REVIEW{Nat. Mater.}{19}{2020}{34}.

\bibitem{Barker2016}
    \Name{Barker J. \and Tretiakov O. A.}
\REVIEW{Phys. Rev. Lett.}{116}{2016}{147203}.

\bibitem{Daniels2019}
    \Name{Daniels M. W., Yu W., Cheng R., Xiao J. \and Xiao D.}
\REVIEW{Phys. Rev. B}{99}{2019}{224433}.

\bibitem{Liang2019}
    \Name{Liang X., \textit{et al.}}
\REVIEW{Phys. Rev. B}{100}{2019}{144439}.

\bibitem{Gong}
\Name{Gong X., Yuan H. Y. \and Wang X. R.}
\REVIEW{Phys. Rev. B}{101}{2020}{064421}.

\bibitem{Yaroslav2002}
  \Name{Tserkovnyak Y., Brataas A. \and Bauer G.E.W.}
  \REVIEW{Phys. Rev. Lett.}{88}{2002}{117601}.

\bibitem{YuanZ2014}
  \Name{Yuan Z., Hals, K.M.D., Liu Y., Starikov A.A., Brataas A. \and Kelly P.J.}
  \REVIEW{Phys. Rev. Lett.}{113}{2014}{266603}.

\bibitem{Cheng2014}
  \Name{Cheng R., Xiao J., Niu Q. \and Brataas  A.}
  \REVIEW{Phys. Rev. Lett.}{113}{2014}{057601}.

\bibitem{Takei2014}
    \Name{Takei S., Halperin B. I., Yacoby A. \and Tserkovnyak Y.}
\REVIEW{Phys. Rev. B}{90}{2014}{094408}.

\bibitem{Kamra2017}
    \Name{Kamra A. \and Belzig W.}
\REVIEW{Phys. Rev. Lett.}{119}{2017}{197201}.

\bibitem{Li2020}
    \Name{Li J. \textit{et al.}}
\REVIEW{Nature}{578}{2020}{70}.

\bibitem{Shen2020}
    \Name{Shen K.}
\REVIEW{Phys. Rev. Lett.}{124}{2020}{077201}.

%\bibitem{Lin2017}
%    \Name{Lin W. \and Chine C. L.}
%\REVIEW{Phys. Rev. Lett.}{118}{2017}{067202}.

\bibitem{Lin2016}
    \Name{Lin W., Chen K., Zhang S. \and Chien C. L.}
\REVIEW{Phys. Rev. Lett.}{118}{2016}{186601}.

\bibitem{Wang2014}
    \Name{Wang H., Du C., Hammel P. \and Yang F.}
\REVIEW{Phys. Rev. Lett.}{113}{2014}{097202}.

\bibitem{Yuan2018}
    \Name{Yuan W. \textit{et al.}}
\REVIEW{Sci. Adv.}{4}{2018}{eaat1098}.

\bibitem{Lebrun2018}
    \Name{Lebrun R. \textit{et al.}}
\REVIEW{Nature}{561}{2018}{222}.

%\cite{Sonin2019}

%\bibitem{Yuan2018}
%    \Name{Yuan W. \textit{et al.}}
%\REVIEW{Sci. Adv.}{4}{2018}{eaat1098}.

\bibitem{Sonin2019}
    \Name{Sonin E. B.}
\REVIEW{Phys. Rev. B}{99}{2019}{104423}.
%\bibitem{Hou2017}
%    \Name{Hou D. \and Chine C. L.}
%\REVIEW{Phys. Rev. Lett.}{118}{2017}{147202}.
%
%\bibitem{Lorenzo2018}
%    \Name{Baldrati L. \textit{et al.}}
%\REVIEW{Phys. Rev. B}{98}{2018}{024422}.
%
%\bibitem{Fischer2018}
%    \Name{Fischer J. \textit{et al.}}
%\REVIEW{Phys. Rev. B}{97}{2018}{014417}.


\bibitem{Hirobe2017}
  \Name{Hirobe D. \textit{et al.}}
 \REVIEW{Nat. Phys.}{13}{2017}{30}.

\bibitem{Camilo2018}
  \Name{Ulloa C. \and Duine R.A.}
 \REVIEW{Phys. Rev. Lett.}{120}{2018}{177202}.


\bibitem{Han2020}
  \Name{Han W., Maekawa S. \and Xie X.-C.}
 \REVIEW{Nat. Mater.}{19}{2020}{139}.

\bibitem{yuanqm}
  \Name{ Yuan H. Y., Yung Man-Hong \and Wang X. R.}
  \REVIEW{Phys. Rev. B}{98}{2017}{060407}.

\bibitem{Wieser2017}
  \Name{ Wieser R.}
  \REVIEW{J. Phys.:Condens. Matter}{29}{2017}{175803}.

\bibitem{Man1972}
  \Name{Manohar C. \and Venkataraman G.}
  \REVIEW{Phys. Rev. B}{5}{1972}{1993}.

\bibitem{Bose1975}
  \Name{Bose S. M., Foo E-Ni \and Zuniga M. A.}
  \REVIEW{Phys. Rev. B}{12}{1975}{3855}.

\bibitem{Xiao2019}
  \Name{Xiao Y. \textit{et al.}}
 \REVIEW{Phys. Rev. B}{99}{2019}{094407}.

\bibitem{yuan2017}
  \Name{Yuan H. Y. \and Wang X. R.}
  \REVIEW{Appl. Phys. Lett.}{110}{2017}{082403}.



\bibitem{yuan2020}
  \Name{Yuan H. Y. \textit{et al.}}
  \REVIEW{Phys. Rev. B}{101}{2020}{014419}.

\bibitem{Parvini2020}
    \Name{Parvini T.S., Bittencourt V. A. S. V. \and Kusminskiy S. V.}
    \REVIEW{Phys. Rev. Research}{2}{2020}{022027}.


\bibitem{Kamra2019}
  \Name{Kamra A. \textit{et al.}}
  \REVIEW{Phys. Rev. B}{100}{2019}{174407}.

\bibitem{Mou2020}
  \Name{Mousolou V. A. \textit{et al.}}
  \REVIEW{arXiv:}{}{2020}{2006.03479}.



\bibitem{Merg2017}
  \Name{Mergenthaler M. \textit{et al.}}
  \REVIEW{Phys. Rev. Lett.}{119}{2017}{147701}.

\bibitem{Everts2020}
  \Name{Everts J. R. \textit{et al.}}
  \REVIEW{Phys. Rev. B}{101}{2020}{214414}.

\bibitem{Gris2018}
  \Name{Grishunin K. \textit{et al.}}
  \REVIEW{ACS Photonics}{5}{2018}{1375}.

\bibitem{Shi2020}
  \Name{Shi L.Y. \textit{et al.}}
  \REVIEW{arXiv:}{}{2020}{2004.05823}.

\bibitem{Johansen2018}
    \Name{Johansen {\O}. \and Brataas A.}
\REVIEW{Phys. Rev. Lett.}{121}{2018}{087204}.

%\bibitem{Pirm2018}
%    \Name{Pirmoradian F., Rameshti B. Z., Miri M. \and Saeidian S.}
%\REVIEW{Phys. Rev. B}{98}{2018}{224409}.
\bibitem{Ra2017}
    \Name{EI-Ganainy R. \textit{et al.}}
\REVIEW{Nat. Phys.}{14}{2018}{11}.


\end{thebibliography}
\end{document}